\documentclass{FBSsuppl}
\newcommand{\btb}{\beta_2}
\newcommand{\bta}{\beta_1}
\newcommand{\atil}{\tilde{a}}
\newcommand{\btil}{\tilde{b}}
\newcommand{\slsh}[1]{\slash \!\!\! #1}

\title{The Off-Shell Nucleon-Nucleon Amplitude: Why it is Unmeasurable in
Nucleon-Nucleon Bremsstrahlung}
\author{Harold W. Fearing}
\institute{TRIUMF, 4004 Wesbrook Mall, Vancouver, B.C., 
Canada V6T 2A3}

\begin{document}
\maketitle
\begin{abstract}
Nucleon-nucleon bremsstrahlung has long been considered a way of getting
information about the off-shell nucleon-nucleon amplitude which would allow one
to distinguish among nucleon-nucleon potentials based on their off-shell
properties. There have been many calculations and many experiments devoted to
this aim. We show here, in contrast to this standard view, that such off-shell
amplitudes are not measurable as a matter of principle. This follows formally
from the invariance of the S-matrix under transformations of the fields. This
result is discussed here and illustrated via two simple models, one applying to
spin zero, and one to spin one half, processes. The latter model is very
closely related to phenomenological models which have been used to study
off-shell effects at electromagnetic vertices.

\end{abstract}

\vspace{-14.5cm}
{\hspace*{\fill} TRI-PP-99-36\\
\hspace*{\fill} (November 15, 1999)}
\vspace{13.75cm}

Almost exactly fifty years ago Ashkin and Marshak \cite{Ashkin49} proposed
using nucleon-nucleon bremsstrahlung to get information about the off-shell
parts of the nucleon-nucleon amplitude. Since then there have been many
calculations, mostly in non-relativistic potential models, of this process and
a number of experiments.  More recently a number of new experiments have been
designed with the express purpose of getting further off shell and thus
providing a more sensitive test of this picture.

The central point of this talk is to show that this historical motivation for
nucleon-nucleon bremsstrahlung is incorrect. In actual fact the off-shell
amplitude is not a physically well defined quantity and {\it as a matter of
principle} cannot be measured in nucleon-nucleon bremsstrahlung, or any other
process.

Let us begin with some general background. The process of interest is $p+p
\rightarrow p + p + \gamma$. Since this process involves an intermediate
virtual particle, it in principle provides access to an off-shell amplitude. It
is interesting because the nucleon-nucleon interaction is perhaps the most
fundamental and certainly most accessible of the strong interactions. Modern
potentials describe the elastic scattering well but in principle produce
different off-shell matrix elements. One can not measure these directly, but
must have some other process to put the particle back on shell. The well known
electromagnetic interaction is the obvious choice.  Thus in this normal, but
incorrect, scenario, measurement of $p+p \rightarrow p + p + \gamma$ in
kinematic regions far enough off shell will determine the half off-shell
T-matrix and distinguish among potentials.

To understand the problem with this approach we first review the standard
potential model description of bremsstrahlung. First observe that in an
abstract mathematical sense the off-shell amplitude is well defined as the
solution of the Lippman-Schwinger equation for off-shell momenta. It is only
part of the full physical bremsstrahlung process however. To get the full
process standard potential models include first the so called external
radiation graphs in which a photon is emitted from an external leg. These
involve off-shell effects both at the strong vertex and at the electromagnetic
vertex. Usually the double scattering term in which the photon is emitted from
a line in between two full scattering T-matrices is also included.  Finally a
few 'contact' graphs are included in an ad hoc fashion. These may include
diagrams with $\Delta$'s or the one with a $\pi \rho \gamma$ or $\pi \omega
\gamma$ vertex. Modern nucleon-nucleon potentials are built from diagrams with
multi meson exchanges, and in principle diagrams with photons attached to any
interior charged line should also be included. There are many such 'contact'
diagrams which are irreducible and thus can not be included in the double
scattering or external radiation graphs. In potential models, these are just
dropped.

At the very simplest, qualitative level one can begin to understand the
ambiguities inherent in the off-shell amplitude as follows.  The amplitude for
a typical bremsstrahlung external radiation graph can be written as
\begin{displaymath}
\left\{ T_{on} + (p^2-m^2)T_{off} \right\} \quad \frac{i}{p^2-m^2} \quad
\Gamma_{em}
\end{displaymath}
corresponding to an on-shell part and an off-shell part of the nucleon-nucleon
interaction, together with propagator and electromagnetic interaction. However,
since the factor $p^2-m^2$ appearing in the off-shell part exactly cancels the
same factor appearing in the propagator, the amplitude can also be written as
\begin{displaymath}
T_{on} \quad \frac{i}{p^2-m^2} \quad  \Gamma_{em} \quad + 
\quad T_{off} i \Gamma_{em}
\end{displaymath}
which has the form of an external radiation graph with on-shell amplitude plus
a contact term. Thus the off-shell terms can be written alternatively as
contact terms, and there will always be an ambiguity in how one separates from
the full measurable process the part associated with an off-shell amplitude.

At a more fundamental level this ambiguity can be related to a theorem
\cite{Haag58} which tells us that a transformation can be made on the fields
appearing in a Lagrangian without changing the S-matrix elements, which
correspond to the physical, measurable quantities. Such transformations
generally change the 'equation of motion' terms in the Lagrangian, which are
those leading to the $p^2-m^2$ factor which gives the off-shell part of the
vertex functions.\cite{Weinberg} This means that such transformations can be
used to change the off-shell T-matrix for the elastic process in an arbitrary
fashion, without changing the full bremsstrahlung amplitude. Thus one can never
measure an off-shell amplitude since the measurable quantity, the
bremsstrahlung cross section, corresponds to an infinite number of different
off-shell amplitudes.

We can summarize what we have learned so far as follows. There will always be
some ambiguity as to what is called off-shell and what is called contact. At a
more formal level field transformations can change the coefficient of the
off-shell amplitude without changing the measurable bremsstrahlung. Thus the
off-shell amplitude is not measurable as there are an infinite number of
different such amplitudes which give the same bremsstrahlung amplitude.

Why wasn't this noticed in the fifty years of work on bremsstrahlung? Perhaps
it was because potential model calculations tend to drop the contact terms from
the beginning and because field transformation concepts, though well known in
some areas of physics, are quite different from the techniques used in most non
relativistic potential model approaches.

We turn now to a couple of simple field theory models which will illustrate the
concepts so far described only in a qualitative way. Consider first a model of
the spin zero bremsstrahlung process, e.g. $\pi^+ + \pi^0 \rightarrow \pi^+ +
\pi^0 + \gamma$, described in detail in ref. \cite{Fearing98}. We take as the
Lagrangian for this process, ${\cal L}={\cal L}_2+{\cal L}_4^{GL}+ \Delta{\cal
L}_4$, where ${\cal L}_2+{\cal L}_4^{GL}$ is the usual chiral perturbation
theory Lagrangian to ${\cal O}(p^4)$ \cite{Gasser84} written in terms of
$\chi$, $U$, which contain the fields, and of the covariant derivative
$D_{\mu}$.  The last term involves the equation of motion and is given
by the expression \cite{Rudy94}
\begin{displaymath}
\Delta{\cal L}_4 =
\bta Tr\left({\cal O}{\cal O}^{\dagger}\right)
+\btb Tr\left[(\chi U^{\dagger}-U\chi^{\dagger}){\cal O}\right].
\end{displaymath}
Here ${\cal O}=0$ is the equation of motion, with ${\cal O}$ a somewhat
complicated function also of $\chi$, $U$, and $D_{\mu}$

This Lagrangian generates a simple effective field theory which allows exact
calculations to a given order. Though it was originally motivated by chiral
perturbation theory, the result to be obtained from it has absolutely nothing
to do with chiral perturbation theory.

Using this Lagrangian we can calculate the elastic amplitude for the process
$\pi^+(p_1) + \pi^0(p_2) \rightarrow \pi^+(p_3) + \pi^0(p_4)$, obtaining a
result given (schematically) as
\begin{displaymath}
\Gamma_{4\pi} = T_{on} + (....)(\Lambda_1+\Lambda_3)+
(....) \btb (\Lambda_1+\Lambda_3) + 
(....) \bta \Lambda_1+(....) \bta \Lambda_3
\end{displaymath}
where $(....)$ are known factors and $\Lambda_i = p_i^2-m_\pi^2$. This is the
exact analog of the off-shell elastic amplitude one would calculate from a
potential by solving the Lippman-Schwinger equation. The off-shell part is
proportional to $\Lambda_1$ or $\Lambda_3$ and for the most part to the
parameters $\bta$ and $\btb$. The on-shell amplitude is obtained by taking the
limit $\Lambda_1,\Lambda_3 \rightarrow 0$

Now what happens when we apply a field transformation to the fields in this
Lagrangian?  To that end consider the transformation $U^\prime \rightarrow {\rm
exp}(iS) U$, for arbitrary $\alpha_1, \alpha_2$, where $S$ is given by
\begin{displaymath}
S=\frac{4i}{F_0^2}\left[\alpha_1{\cal O}-
\alpha_2\left(\chi U^{\dagger}-U \chi^{\dagger}-
\frac{1}{2} Tr(\chi U^{\dagger}-U\chi^{\dagger})\right)\right].
\end{displaymath}
Under this transformation ${\cal L}_2(U^\prime) \rightarrow {\cal L}_2(U) +
\delta{\cal L}_2(U)$ with
\begin{displaymath}
\delta{\cal L}_2 =\alpha_1 Tr({\cal O}{\cal O}^{\dagger})
+\alpha_2Tr((\chi U^\dagger-U\chi^\dagger){\cal O})  
\end{displaymath}
Clearly $\delta{\cal L}_2(U) \sim \Delta{\cal L}_4$. Thus what the
transformation does, since $\alpha_1, \alpha_2$ are arbitrary, is to
arbitrarily change the values of the coefficients $\bta,\btb$ in the
Lagrangian, i.e. to change the coefficients which give the off-shell elastic
amplitude. By the general result \cite{Haag58} such transformation does not
change the full measurable amplitude, though it clearly changes the off-shell
part of the elastic amplitude.
                   
Now let us apply this model to bremsstrahlung. The external radiation diagrams
can be obtained from the off-shell elastic amplitude, a propagator, and the
electromagnetic $\pi\pi\gamma$ vertex, all consistently calculated using the
model Lagrangian.  The result is \cite{Fearing98} in abbreviated form,
\begin{eqnarray*}
M_3+M_1 &=& T_{on} (\frac{\epsilon \cdot p_3}{k \cdot p_3}-
\frac{\epsilon \cdot p_1}{k \cdot p_1})+{\rm constant} \\
& +& \bta  (...{\rm A}...) + \btb (...{\rm B}...)
\end{eqnarray*}
where the A and B pieces are known kinematic dependent factors.

$M_3+M_1$ is the analogue of what would be calculated in a potential model from
the half off-shell T-matrix, a non-relativistic propagator, and an
electromagnetic vertex. In a potential model it would be compared with data to
extract the off-shell information 'contained' in $\bta$ and $\btb$. But, this
approach has to be wrong because field transformations change the value of
$\bta$ and $\btb$ arbitrarily without changing the value of the bremsstrahlung
amplitude. In fact the bremsstrahlung amplitude must be independent of $\bta$
and $\btb$ since 0 is one possible value.

To understand what is happening, consider the contact term, which can be
calculated explicitly in this model, unlike in potential model calculations.
\begin{eqnarray*}
\Gamma_{4\pi\gamma}&=& {\rm const.} - 
\bta  (...{\rm A}...) - \btb (...{\rm B}...)
\end{eqnarray*} 
Comparison of the contact term with $M_3+M_1$ shows that all terms involving
the parameters $\bta, \btb$, which govern the strength of the off-shell
amplitude, cancel in the full result for bremsstrahlung which is thus
independent of $\bta$ and $\btb$ as it must be. A similar result is obtained
when this Lagrangian is used for Compton scattering.\cite{Scherer95}

The fragment of the amplitude, $M_1+M_3$, which is analogous to the usual
potential model bremsstrahlung result, does however depend on $\bta,\btb$ and
if one considers it alone, one is led to the (spurious) conclusion that
off-shell amplitudes can be measured by measuring bremsstrahlung.

Let us now consider a different model \cite{Fearing99}, one applicable to spin
one-half particles. This model corresponds very closely to the types of
phenomenological models used in calculations investigating off-shell effects at
electromagnetic vertices, e.g. \cite{Nyman70}. The prototype reaction will be
$p + n \rightarrow p + n +\gamma$, i.e. proton-neutron bremsstrahlung. We
consider the p-n system, rather than the p-p system just to reduce the number
of diagrams to be considered and to avoid the extra algebra required for
identical particles.

We thus start with the Lagrangian,
\begin{equation}
{\cal L}_0 = \overline{\Psi}(i \slsh{D} -m) \Psi -\frac{e \kappa}{4 m}
\overline{\Psi} \sigma_{\mu \nu} F^{\mu \nu} \Psi +g \overline{\Psi} \Psi
\overline{\Phi} \Phi.
\end{equation}
Here $D$ is the covariant derivative, $e$ and $\kappa$ are the proton charge
and anomalous magnetic moment, $A_\mu$, $F_{\mu\nu}$ are photon field and field
strength tensor, and $\Psi \sim$ proton, $\Phi \sim$ neutron.  This Lagrangian
leads to the standard electromagnetic coupling to a spin one-half particle, the
standard free equation for the proton and to a standard, but simplified, scalar
boson exchange form for the strong interaction.  However it generates no
off-shell effects to lowest order.

Now consider a field transformation on this Lagrangian of the form
\begin{displaymath}
\Psi^\prime \rightarrow \Psi + \atil g \overline{\Phi}\Phi\Psi
+\btil e \sigma_{\mu\nu} F^{\mu\nu} \Psi
\end{displaymath}
where $\atil$ and $\btil$ are arbitrary constants This transformation generates
a new Lagrangian
\begin{displaymath}
{\cal L}_0(\Psi^\prime) = {\cal L}_0(\Psi) + \Delta_1{\cal L}(\Psi) +
\Delta_2{\cal L}(\Psi)
\end{displaymath}
The new piece of the Lagrangian $\Delta_1{\cal L}(\Psi)$ generates an off-shell
contribution to the strong amplitude proportional to $\atil$ and an off-shell
part of the electromagnetic vertex proportional to $\btil$ as well as a contact
term necessary for gauge invariance. $\Delta_2{\cal L}(\Psi)$ generates only
contact terms.

If we now take ${\cal L}_0(\Psi) + \Delta_1{\cal L}(\Psi)$ as our Lagrangian we
have something corresponding very closely to the Lagrangian used in
phenomenological calculations such as those of \cite{Nyman70,Kondratyuk}. It
produces off-shell contributions at both strong and electromagnetic
vertices. One could compare the predictions of this Lagrangian for
bremsstrahlung with data and in principle extract values of the parameters
$\atil$ and $\btil$. However this does not give any information about off-shell
nucleon-nucleon amplitudes. Since the change in the Lagrangian, $\Delta_1{\cal
L}(\Psi) + \Delta_2{\cal L}(\Psi)$, originates in a field transformation it can
not affect the bremsstrahlung amplitude. This means that the amplitude using
the full Lagrangian must be independent of the parameters $\atil$ and $\btil$,
which can be verified explicitly. As a corollary, the Lagrangian ${\cal
L}_0(\Psi) - \Delta_2{\cal L}(\Psi)$ must give exactly the same bremsstrahlung
result as ${\cal L}_0(\Psi) + \Delta_1{\cal L}(\Psi)$. But the former
Lagrangian contains only contact terms and no off-shell effects while the
latter has mainly off-shell effects. One could also use any linear combination
of these two Lagrangians. Thus there are an infinite number of Lagrangians,
with different proportions of off-shell effects and contact terms which give
the same bremsstrahlung amplitude. So again we see that the concept of an
off-shell amplitude as a part of a bremsstrahlung amplitude is not well defined
and such off-shell amplitudes are not measurable.

Let us now summarize what has been learned.  Field transformations change the
coefficients of the off-shell part of the elastic amplitude without changing
the bremsstrahlung amplitude.  In effect such transformations simply move terms
back and forth between the bremsstrahlung diagrams containing off-shell
amplitudes and the contact terms.  This means that the off-shell contributions
are not uniquely defined and therefore that the 'off-shell amplitude' is not
measurable in bremsstrahlung reactions, contrary to historical
expectations. Hence the claim that measuring bremsstrahlung will distinguish
among potentials on the basis of their off-shell behavior is just not valid.

This result means that most previous calculations and experiments dealing with
nucleon-nucleon bremsstrahlung were driven by an aim -- to measure off-shell
effects -- which is just not possible. Does this make nucleon-nucleon
bremsstrahlung any less interesting? The answer is clearly no.  What emerges
from these results is the importance of the contact terms.  These contact terms
originate from the photon probe of the currents internal to the strong
interaction. To understand them in detail one must understand these
interactions at a microscopic level, which is probably much more interesting
and gives us much more insight about the physical mechanisms involved, than
does understanding gross properties of a phenomenological potential.

 This means however that for the future we need both comprehensive experiments
with enough data over a wide enough kinematic range to distinguish details of
the process and also microscopic calculations which include specifically
details of the contact terms.

Finally nothing here depends specifically on the bremsstrahlung process, and so
it is probable that off-shell amplitudes are unmeasurable in any process. Thus
calculations purporting to show sensitivity to off-shell amplitudes should be
viewed with suspicion.

The author would like to thank Stefan Scherer who has collaborated on much of
what is reported here. The work was also supported in part by a grant from the
Natural Sciences and Engineering Research Council of Canada.


\frenchspacing
\begin{thebibliography}{99}

\bibitem{Ashkin49} J.~Ashkin and R.~Marshak, Phys. Rev. {\bf 76}, 58 (1949);
Erratum, {\em ibid}, 989.

\bibitem{Haag58} R. Haag, Phys. Rev. {\bf 112}, 669 (1958);
J. S. R. Chisholm, Nucl. Phys. {\bf 26}, 469 (1961); 
S. Kamefuchi, L. O'Raifeartaigh, and A. Salam,
Nucl. Phys. {\bf 28}, 529 (1961).

\bibitem{Weinberg} S. Weinberg {\em Quantum Theory of Fields }, vol. 1, p. 
331 (Cambridge Univ. Press, 1995).

\bibitem{Fearing98} H.~W.~Fearing, Phys. Rev. Lett. {\bf 81}, 758 (1998).

\bibitem{Gasser84} J. Gasser and H. Leutwyler, Ann. Phys. {\bf 158}, 
142 (1984).

\bibitem{Rudy94} T. E. Rudy, H. W. Fearing, and S. Scherer, Phys. Rev. C
{\bf 50}, 447 (1994).

\bibitem{Scherer95} S. Scherer and H. W. Fearing, Phys. Rev. C {\bf 51}, 359
(1995).

\bibitem{Fearing99} H. W. Fearing and S. Scherer, TRI-PP-99-30, 
nucl-th/9909076.

\bibitem{Nyman70} E. M. Nyman, Nucl. Phys. {\bf A154}, 97 (1970);
{\bf A160}, 517 (1971).

\bibitem{Kondratyuk} S. Kondratyuk, G. Martinus, and O. Scholten, Phys.
Lett. B {\bf 418}, 20 (1998).

\end{thebibliography}
\end{document}